\documentclass[%
 preprint,
 superscriptaddress,
 amsmath,amssymb,
 aps,
 showkeys,
 titlepage,
 endfloats*.   %move all floats to the end of the document and put each on a separate page
]{revtex4-2}

\usepackage[utf8]{inputenc}
\usepackage[english]{babel}

\usepackage{graphicx}% Include figure files
\usepackage{dcolumn}% Align table columns on decimal point
\usepackage{bm}% bold math

\usepackage[colorlinks = true,
            linkcolor = blue,
            urlcolor  = blue,
            citecolor = red,
            anchorcolor = blue]{hyperref}

\begin{document}

%\preprint{APS/123-QED}

\title{Coupled electron-phonon hydrodynamics in two-dimensional semiconductors} 

\author{Yujie Quan}
\affiliation{Department of Mechanical Engineering, University of California, Santa Barbara, CA 93106, USA}

\author{Bolin Liao}
\email{bliao@ucsb.edu} \affiliation{Department of Mechanical Engineering, University of California, Santa Barbara, CA 93106, USA}

\date{\today}

\begin{abstract}
Electronic and thermal transport properties in two-dimensional (2D) semiconductors have been extensively investigated due to their potential to miniaturize transistors.
Microscopically, electron-phonon interactions are considered the dominant momentum relaxation mechanism for electrons that limits carrier mobility beyond cryogenic temperatures.
However, when electrons and phonons are considered as a single system, electron-phonon interactions conserve the total momentum and energy, leading to the possibility of low-dissipation transport.
In this work, we systematically investigate the momentum circulation between electrons and phonons and its impact on carrier transport properties in 2D semiconductors given their strong electron-phonon interactions.
We find that, when momentum circulation is taken into account, the total momentum in the coupled electron-phonon system is weakly dissipated, leading to a coupled electron-phonon hydrodynamic transport regime, in which electrons and phonons exhibit a joint drift motion rather than separate diffusive behaviors.
In this new transport regime, charge transport properties are significantly enhanced.
Contrary to previous belief, our results demonstrate that low-dissipation charge transport can occur despite strong electron-phonon interactions when there is effective momentum circulation between electrons and phonons mediated by the strong interactions.
Our work advances fundamental understandings of carrier transport in 2D semiconductors.

\end{abstract}

%Use showkeys class option if keyword display desired
                            
\maketitle

%\tableofcontents

\section{Introduction}
\begin{figure}[!htb]
\includegraphics[width=1\textwidth]{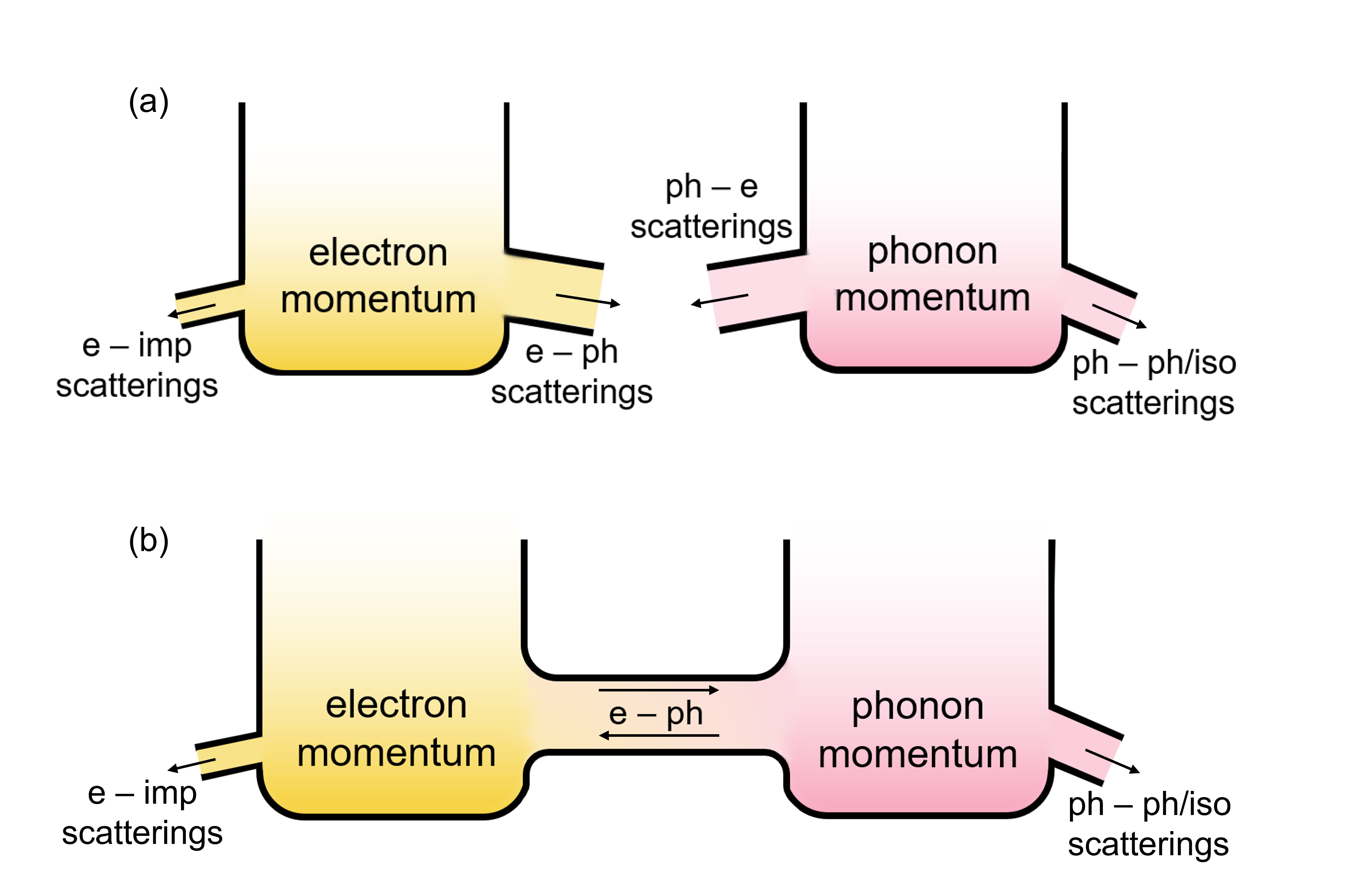}
\caption{\textbf{A schematic of the concept of low-dissipation conduction despite strong electron-phonon interactions.} (a) A schematic of momentum dissipation channels in electron and phonon systems when treated separately. (b) A schematic of momentum dissipation channels in a coupled electron-phonon system. The momentum in electron and phonon systems circulates through electron-phonon interactions, which no longer cause momentum dissipation. e-ph: electron-phonon; ph-e: phonon-electron; e-imp: electron-impurity; ph-ph: phonon-phonon; ph-iso: phonon-isotope.}
\label{fig:fig1}
\end{figure}

%Two-dimensional (2D) semiconductors have gained extensive interest due to their promise of realizing miniaturization in transistor size~\cite{chhowalla2016two,das2021transistors}. 
%As a result, extensive theoretical research has been conducted to better understand their transport properties ~\cite{mir2020recent,gu2018colloquium,cheng2020two}, which are essential in designing and optimizing the performance of 2D semiconductor-based devices. 
%For example, carrier mobility has been widely studied in graphene, 2D transition metal dichalcogenides, 2D phosphorous, and other 2D semiconductors~\cite{mir2020recent}, and their generally low electron mobility compared with three-dimensional (3D) semiconductors is attributed to high electron-phonon scattering phase space~\cite{cheng2020two}, which characterizes the number of electron-phonon pairs that satisfy both momentum and energy conservation laws.
%In addition to electronic transport properties, thermal transport properties are also extensively studied in these materials~\cite{gu2018colloquium}. 
%The underlying mechanism that governs both electronic and thermal transport properties is related to microscopic scattering processes.
Electron-phonon interactions are central to understanding the fundamental properties and behaviors of both metals and semiconductors. 
For instance, Cooper pairs are formed due to the attractive interaction mediated by phonons, leading to superconductivity~\cite{bardeen1957theory}.
Furthermore, electron-phonon interactions play a crucial role in driving charge density wave transitions~\cite{zhu2015classification,zhu2017misconceptions}.
In terms of transport properties in semiconductors and metals, electron-phonon interactions are usually considered the primary energy and momentum sink in the electron system, while phonon-electron scatterings can also lead to momentum relaxation within the phonon system~\cite{liao2015significant,quan2021impact}, as shown in Fig.~\ref{fig:fig1}(a). This consideration of electrons and phonons as separate systems has contributed to the impression that strong electron-phonon interactions often lead to highly dissipative transport with low charge mobility. 
However, in reality, electrons and phonons in the same material are coupled and should be treated collectively as a single system during charge and heat transport. From this point of view, despite frequent momentum and energy exchanges through strong electron-phonon interactions, the overall momentum and energy of the coupled system are conserved, as shown in Fig.~\ref{fig:fig1}(b), giving rise to a coupled electron-phonon hydrodynamic transport regime. In this regime, contrary to the conventional belief, strong electron-phonon interactions can actually lead to low-dissipation, or even dissipationless, transport. 
% Until now, this coupled electron-phonon hydrodynamic transport has been primarily explored experimentally in metallic and semimetallic systems, and the associated theory has been limited to simplified models ~\cite{huang2021electron,levchenko2020transport,fritz2024hydrodynamic}.
% Given the strong electron-phonon interactions in 2D semiconductors due to the large electron-phonon phase space~\cite{cheng2020two}, it is crucial to revisit the transport properties from the perspective of coupled electron-phonon transport.

Hydrodynamic transport of fundamental microscopic (quasi)particles shares similarities with the macroscopic transport of fluids, where the momentum is conserved, and the fluid elements move collectively.
In the case of electron flow, it has been predicted that electronic transport enters the hydrodynamic regime when momentum-conserving electron-electron scatterings dominate over momentum-relaxing scatterings with phonons, impurities, and boundaries~\cite{crossno2016observation,bandurin2016negative,moll2016evidence,varnavides2023charge}. 
Thanks to advancements in high-quality materials growth and precise transport measurements, distinctive electronic transport properties arising from electron hydrodynamics, such as the non-monotonic behavior of differential resistance with respect to lattice temperatures and the violation of Wiedemann-Franz law, have now been observed in several material systems, including  2D electron gases~\cite{de1995hydrodynamic}, graphene~\cite{crossno2016observation}, and Weyl semimetals~\cite{gooth2018thermal}.
Similarly, the dominance of momentum-conserving normal phonon-phonon scatterings over umklapp phonon-phonon scatterings leads to phonon hydrodynamics~\cite{lee2015hydrodynamic,cepellotti2015phonon}, where phonons show macroscopic drift motion. 
In the phonon hydrodynamic regime, the response to a perturbation, such as a heat pulse, differs significantly from the diffusive and ballistic regimes. 
The propagation of the induced temperature wave because of the heat pulse, referred to as second sound, can occur over distances longer than the ballistic transport limit~\cite{lee2015hydrodynamic}.
In contrast, in the diffusive regime, the heat pulse is heavily damped and unable to propagate due to the dominance of umklapp scatterings. 
The second sound in solids was first observed in 3D materials at cryogenic temperatures~\cite{ackerman1966second,narayanamurti1972observation,jackson1970second}, and recent experiments have directly observed the second sound in graphite above 100~K~\cite{huberman2019observation,ding2022observation}. 
In both electron and phonon hydrodynamics, electron-phonon scatterings are generally considered destructive due to momentum dissipation.
This conclusion is based on the assumption that phonons remain in equilibrium when electrons lose momentum by scattering with phonons, and vice versa.
However, the nonequilibrium electrons (phonons) can drive phonons (electrons) away from their equilibrium states, and the momentum exchange between nonequilibrium electrons and phonons creates a momentum circulation between the two systems, typically known as electron-phonon drag~\cite{gurevich1989electron,herring1954theory}.
The electron-phonon drag effect has been found to significantly increase transport properties, even at room temperature, including thermopower~\cite{zhou2015ab,protik2020electron,li2023high}, carrier mobility~\cite{quan2023significant}, and thermal conductivity~\cite{quan2024electron}. 
However, another consequence of the strong momentum circulation\textemdash non-dissipative electron-phonon scatterings that result in coupled electron-phonon hydrodynamic transport\textemdash has not been fully explored.

While the theory of phonon drag effect on electronic transport properties was developed in the last century, the theory of coupled electron-phonon hydrodynamics has only emerged in recent years ~\cite{huang2021electron,levchenko2020transport}.
The coupled Boltzmann transport equations (BTE) were solved with approximations of collision integrals and simplified models, from which the electrical and thermal conductivities are predicted to be higher~\cite{levchenko2020transport}. 
In addition, the electron-phonon hydrodynamics were discussed under different temperature regimes with distinct thermodynamic properties and hydrodynamic transport coefficients~\cite{huang2021electron}.
In an earlier study, electron-phonon interactions were found to contribute to hydrodynamic electronic transport in $\rm WP_2$, but the perspective on phonon transport is lacking~\cite{coulter2018microscopic}.
Experimentally, evidence of coupled electron-phonon hydrodynamics was observed in $\rm NbGe_2$~\cite{yang2021evidence}, characterized by strong electron-phonon interactions and the strong suppression of momentum-relaxing phonon-phonon scatterings.
Nevertheless, previous studies have been limited to (semi)metal systems and the definite characteristic of the coupled electron-phonon hydrodynamics\textemdash the collective drift motion of electrons and phonons\textemdash has not been examined. 
Given that 2D semiconductors typically exhibit strong electron-phonon interactions, which are considered the dominant momentum sink that limits transport coefficients, it is important to investigate the potential existence of coupled electron-phonon hydrodynamics in those materials, which could lead to significantly enhanced transport coefficients that are beneficial for future device design.

In this work, we investigate the coupled electron-phonon hydrodynamics in 2D semiconductors from first-principles by solving the fully coupled electron-phonon BTEs~\cite{protik2022elphbolt}. 
$\rm MoS_2$, which was found to have strong electron-phonon interactions due to its polar nature without inversion symmetry, is chosen as an example.
We also compare the results with 2D black phosphorene, in which the electron-phonon interactions are weaker.
Surprisingly, we find that with strong electron-phonon interactions, electrons and phonons in $\rm MoS_2$ follow a collective drift motion with the same drift velocity in response to either a temperature gradient or an electric field\textemdash a signature of coupled electron-phonon hydrodynamic transport\textemdash at and above 100~K, which is much higher than temperatures reported in previous experiments for hydrodynamic electron transport. 
Furthermore, we observe a significant increase in the calculated transport coefficients in the coupled electron-phonon hydrodynamic regime in $\rm MoS_2$ compared to the case where momentum circulation is neglected, and this increase greatly exceeds the enhancement seen in black phosphorene, where only the electron-phonon drag effect is present and no coupled electron-phonon hydrodynamics occur.
This strong enhancement is attributed to the low dissipation of total momentum in electron and phonon systems, despite strong interactions between them. 
Our work demonstrates that electrical and thermal conductivities can remain high despite strong electron-phonon interactions and that the strength of these interactions is not the determinant factor that limits electrical mobility, which are important findings that have been previously overlooked. Our result also suggests that low dissipation transport can be realized in materials with strong electron-phonon coupling but minimal momentum-loss channels such as phonon-phonon umklapp scatterings and impurity scatterings.
%In general: please start a newline for each sentence: this helps locate accurate positions in the manuscript using Overleaf.

%Use this format for citation keys: first-author-last-name/year/first-word-of-title. This is the default format for google scholar import.

%Introduction is the most important part of a paper and should be a more detailed version of the abstract. An exemplar organization of the Introduction:

%\textbf{[Paragraph 1: General Background.] }

%What is the material/property/physics/technique that this paper is about? Why is this material/property/physics/technique important (e.g. discuss values in terms of fundamental understanding and/or practical application)? What are the challenges and/or unknowns?
%If needed, another paragraph can be added to provide more specific background knowledge for readers not familiar with the topic.

%\textbf{[Paragraph 2: Context and Previous Work.]}

%What did other people do previously about these challenges/unknowns? What methods/theories were developed? What assumptions were made? What are the remaining challenges/unknowns that can be pursued?

%\textbf{[Paragraph 3: Summary of Contribution.]}

%Usually start this paragraph with ``In this work, we used/applied/developed...''. Clearly state why/how our approach is different from previous approaches and what advancements are enabled by our new approach. Briefly summarize our main findings in this work. Provide one to two sentences on the importance of our main findings in context. 

\section{Results and Discussions}
The macroscopic joint drift motion of electrons and phonons, which is a signature of coupled electron-phonon hydrodynamic transport, is manifested in their nonequilibrium distribution functions. 
As derived in Ref~\cite{levchenko2020transport}, in the coupled electron-phonon hydrodynamic regime, the displaced equilibrium distribution of electrons $f^h$ and phonons $n^h$ can be written as:
\begin{equation} \label{eq:2}
    f^h = \dfrac{1}{\exp ( \frac{\hbar ((\varepsilon - \mu)/\hbar - \mathbf{q} \cdot \mathbf{u}) }{k_{\rm B} T}) + 1}
\end{equation}
\begin{equation} \label{eq:3}
    n^h = \dfrac{1}{\exp ( \frac{\hbar (\omega - \mathbf{q} \cdot \mathbf{u}) }{k_{\rm B} T}) - 1},
\end{equation}
where $\varepsilon$, $\mu$, $\mathbf{q}$, and $\omega$ represent electron energy, chemical potential, wavevector, and phonon frequency, respectively. 
$\mathbf{u}$ is the joint drift velocity for both electrons and phonons. 
Assuming small displacements from equilibrium, these two displaced distributions can be linearized to 
\begin{equation} \label{eq:4}
    f^h = f_0 + \dfrac{\hbar}{k_{\rm B}T}f_0(1-f_0)\mathbf{q}\cdot \mathbf{u}
\end{equation}
\begin{equation} \label{eq:5}
    n^h = n_0 + \dfrac{\hbar}{k_{\rm B}T}n_0(1+n_0)\mathbf{q}\cdot \mathbf{u} .
\end{equation}
According to Eq.~\ref{eq:2}-Eq.~\ref{eq:5}, the joint drift velocity can be easily obtained by plotting the normalized deviation of the distribution from the equilibrium Fermi-Dirac distribution or Bose-Einstein distribution, defined as $(f-f_0)/(f_0(1-f_0))$ and $(n-n_0)/(n_0(1+n_0))$.

\begin{figure}[!htb]
\includegraphics[width=1\textwidth]{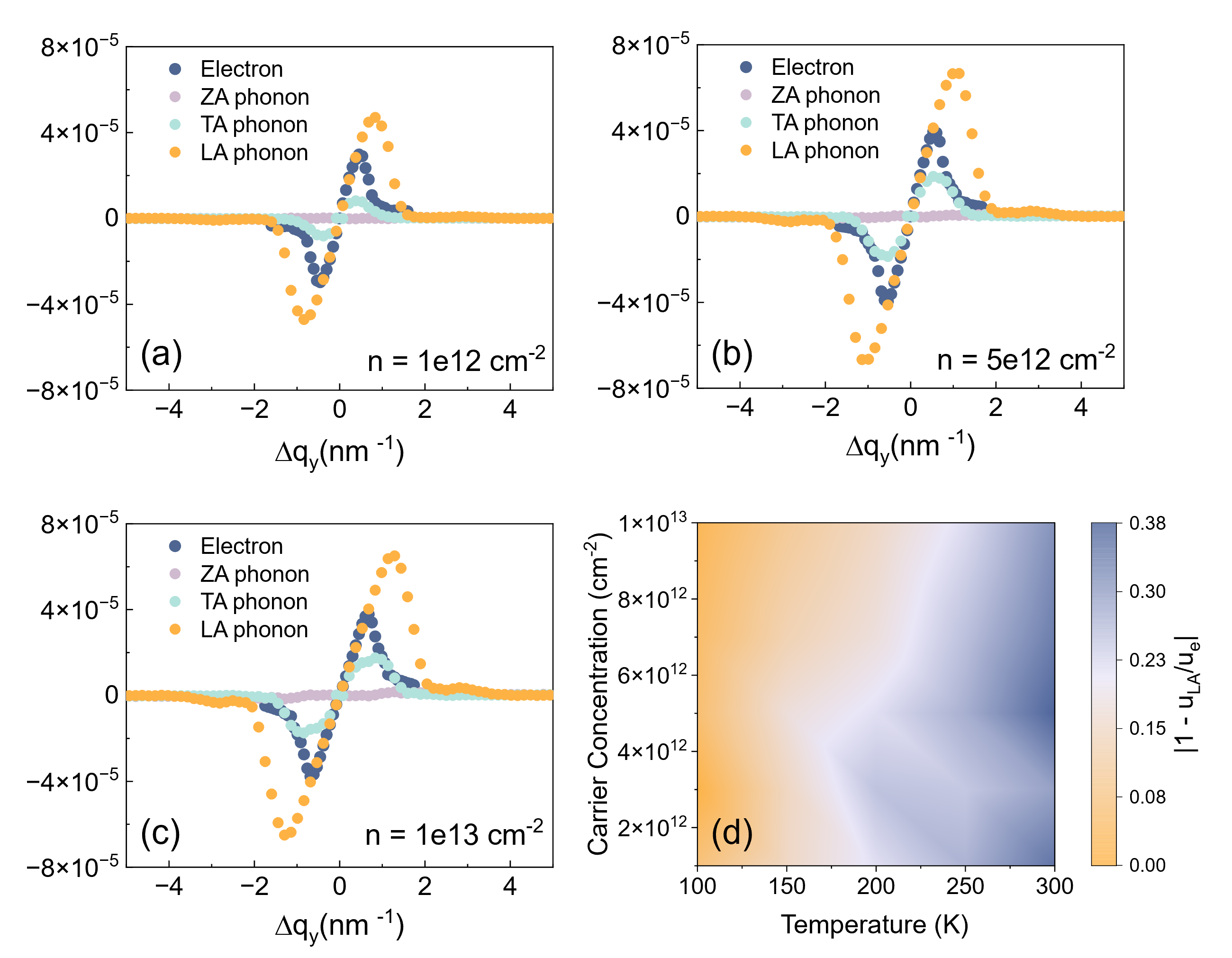}
\caption{\textbf{The normalized deviation of distribution functions of electrons and phonons from equilibrium in $\rm \mathbf{MoS_2}$ in response to a y-direction unit electric field.} (a)-(c) The normalized deviation of distribution at 100~K with different carrier concentrations. The x-axis represents the distance in reciprocal space from the $K$ point for electrons and the $\Gamma$ point for phonons. The linear relationship between the normalized deviation and $\Delta q_y$ for both electrons and phonons demonstrates the hydrodynamic feature. The same slope between the LA phonons and electrons shows the joint drift motion, indicating the coupled electron-phonon hydrodynamics. (d) The deviation of the drift velocity ratio between LA phonons and electrons from unity as a function of temperature and carrier concentration. Higher carrier concentration facilitates coupled electron-phonon hydrodynamics.}
\label{fig:fig2}
\end{figure}

The normalized deviations of the distribution of both electrons and phonons in MoS$_2$ at 100 K in response to a unit electric field are shown in Fig.~\ref{fig:fig2}. 
It is noted that phonons cannot directly couple to the electric field, and their deviation from equilibrium is entirely due to the interactions with nonequilibrium electrons, which drive the phonons away from their equilibrium states. 
The normalized deviations of $\rm MoS_2$ at 100~K with carrier concentration in the range of $10^{12}~\rm cm^{-2}$ and $10^{13}~\rm cm^{-2}$ are shown in Fig.~\ref{fig:fig2}.
In $\rm MoS_2$, the conduction band minimum is located at $K$ point, near which electrons involved in transport deviate from equilibrium.
Phonons, however, deviate from equilibrium near the $\Gamma$ point due to strong electron-phonon interactions occurring in that region of reciprocal space.
For direct comparison, the normalized deviations of the electron and phonon distributions are plotted in the same figure, with the x-axis showing the distance in reciprocal space from the $K$ point for electrons and the $\Gamma$ point for phonons. 
It is clearly seen that the normalized deviation exhibits a linear relationship with the wavevector $\Delta q_y$ along the electric field direction (y-direction), with the slope related to the drift velocity, demonstrating a hydrodynamic feature.
In addition to electrons, the normalized deviations of three acoustic phonon modes are plotted.
It is found that the longitudinal acoustic (LA) mode shares the same drift velocity as the electrons. 
In contrast, the transverse acoustic (TA) mode has a slower drift velocity, and the out-of-plane acoustic (ZA) mode has a negligible response to the electric field. 
This can be explained by the strength of electron-phonon interactions since this is the only driving force that leads to a response to an electric field in the phonon system.
In our previous work, we demonstrated that in $\rm MoS_2$, the interactions between electrons and the LA phonons are particularly strong, while the interactions with the TA phonons are weaker, and they are notably weak with the ZA mode~\cite{quan2024electron}.
The excellent agreement in drift velocity between electrons and LA phonons indicates a joint drift behavior, which is a hallmark of coupled electron-phonon hydrodynamics.
It is important to note that our calculations account for umklapp phonon-phonon scatterings, phonon-isotope scatterings, and electron-impurity scatterings.
Nevertheless, thanks to the strong electron-phonon interactions and momentum circulation, the feature of coupled electron-phonon hydrodynamics is still preserved.
The normalized distribution deviations at higher temperatures in response to a unit electric field are shown in the Supplementary Information.
It is found that even at higher temperatures, the coupled electron-phonon hydrodynamic characteristic still persists.
At higher temperatures, the drift velocity of the LA phonons is slightly lower than that of the electrons, which can be attributed to increased umklapp phonon-phonon scatterings, acting as a momentum sink in phonon systems. 
Fig.~\ref{fig:fig2}(d) illustrates the deviation of the drift velocity ratio between LA phonons and electrons from unity, expressed as $\left | 1 - \dfrac{u_{LA}}{u_e}\right |$, as a function of both temperature and carrier concentration. 
It is shown that even at higher temperatures, electrons and LA phonons tend to share a closer drift velocity at higher carrier concentrations, due to the increased electron density that promotes more frequent interactions, driving phonons away from equilibrium and thus leading to coupled electron-phonon hydrodynamics.
For comparison, the normalized deviation of black phosphorene in response to a unit electric field at 100~K is shown in Fig.~S5.
Although momentum circulation is also considered, there is no joint drift motion between electrons and phonons, which is due to the relatively weak electron-phonon interactions.

\begin{figure}[!htb]
\includegraphics[width=1\textwidth]{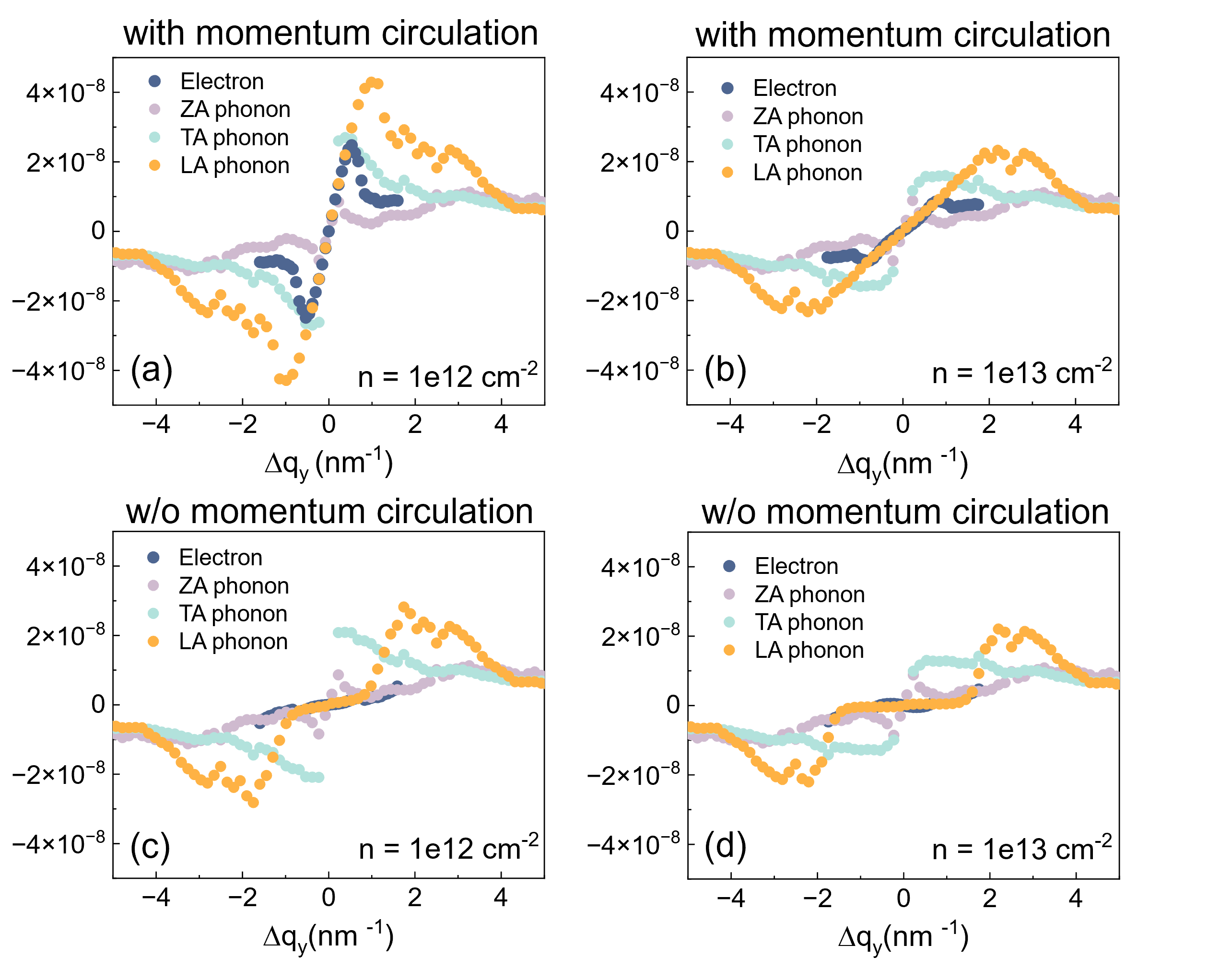}
\caption{\textbf{The normalized deviation of distribution functions of electrons and phonons in $\rm \mathbf{MoS_2}$ from equilibrium in response to a y-direction unit temperature gradient at 100~K.} The top panels show the calculated normalized deviation with momentum circulation, while the bottom panels show the calculated normalized deviation without momentum circulation. When momentum circulation is considered, both electrons and the LA phonons exhibit a linear relationship with $\Delta q_y$ and share the same slope, indicating the coupled electron-phonon hydrodynamics.}
\label{fig:fig3}
\end{figure}

In addition to the response to an electric field, electrons and phonons can also be driven out of equilibrium by a temperature gradient.
The normalized deviations of electrons and phonons in response to a unit temperature gradient are shown in Fig.~\ref{fig:fig3}.
Unlike their response to an electric field, phonons can directly couple with a temperature gradient.
Therefore, the normalized deviations are plotted with and without considering the interactions between nonequilibrium electrons and phonons for comparison. 
The normalized deviations of $\rm MoS_2$ at $n = 10^{12}$~cm$^{-2}$ and $n = 10^{13}$~cm$^{-2}$ at 100~K are shown in the main text, while the normalized deviations at other carrier concentrations and higher temperatures are shown in the Supplementary Information.
The bottom panels show the normalized deviations without considering the momentum circulation between nonequilibrium electrons and phonons.
It is observed that there is no coupling between electrons and phonons, and neither exhibits a hydrodynamic drift behavior.
The non-drift phonons are consistent with previous calculations~\cite{cepellotti2015phonon}, indicating that the phonons are not in the hydrodynamic regime at this temperature.
In contrast, when momentum circulation is taken into account, as shown in the upper panels of Fig.~\ref{fig:fig3}, the normalized deviations of LA phonons and electrons change significantly.
They exhibit drift motion and share the same drift velocity, indicating that they have entered the coupled electron-phonon hydrodynamic regime. 
The normalized deviations of TA phonons and ZA phonons, however, have little change after considering the momentum circulation, which is due to their weak interactions with electrons.
At higher temperatures, due to the increased phonon anharmonicity, strong umklapp phonon-phonon scatterings dominate over electron-phonon interactions, making the drift feature less pronounced.
Nevertheless, even at room temperature, the significant changes in the distribution functions of both electrons and LA phonons due to the momentum circulation still indicate a tendency towards the coupled hydrodynamic transport regime, as shown in Fig.~S10.
Figure~S11 shows the deviation of the drift velocity ratio between LA phonons and electrons from unity in response to a unit temperature gradient as a function of temperature and carrier concentration. Since the normalized deviation of distributions does not follow a strictly linear relationship with the wavevector at higher temperatures, the drift velocity is analyzed only up to 200~K.
Although the drift velocities of electrons and LA phonons show slight deviations at higher temperatures, similar to the behavior observed in response to an electric field, this deviation does not exhibit a clear relationship with carrier concentration.
In contrast, the response to a unit temperature gradient in black phosphorene at 100~K is shown in Fig.~S12. 
It is observed that the distribution of electrons and phonons changes very little with or without momentum circulation, indicating that phosphorene does not exhibit temperature-driven coupled electron-phonon hydrodynamics.

\begin{figure}[!htb]
\includegraphics[width=1\textwidth]{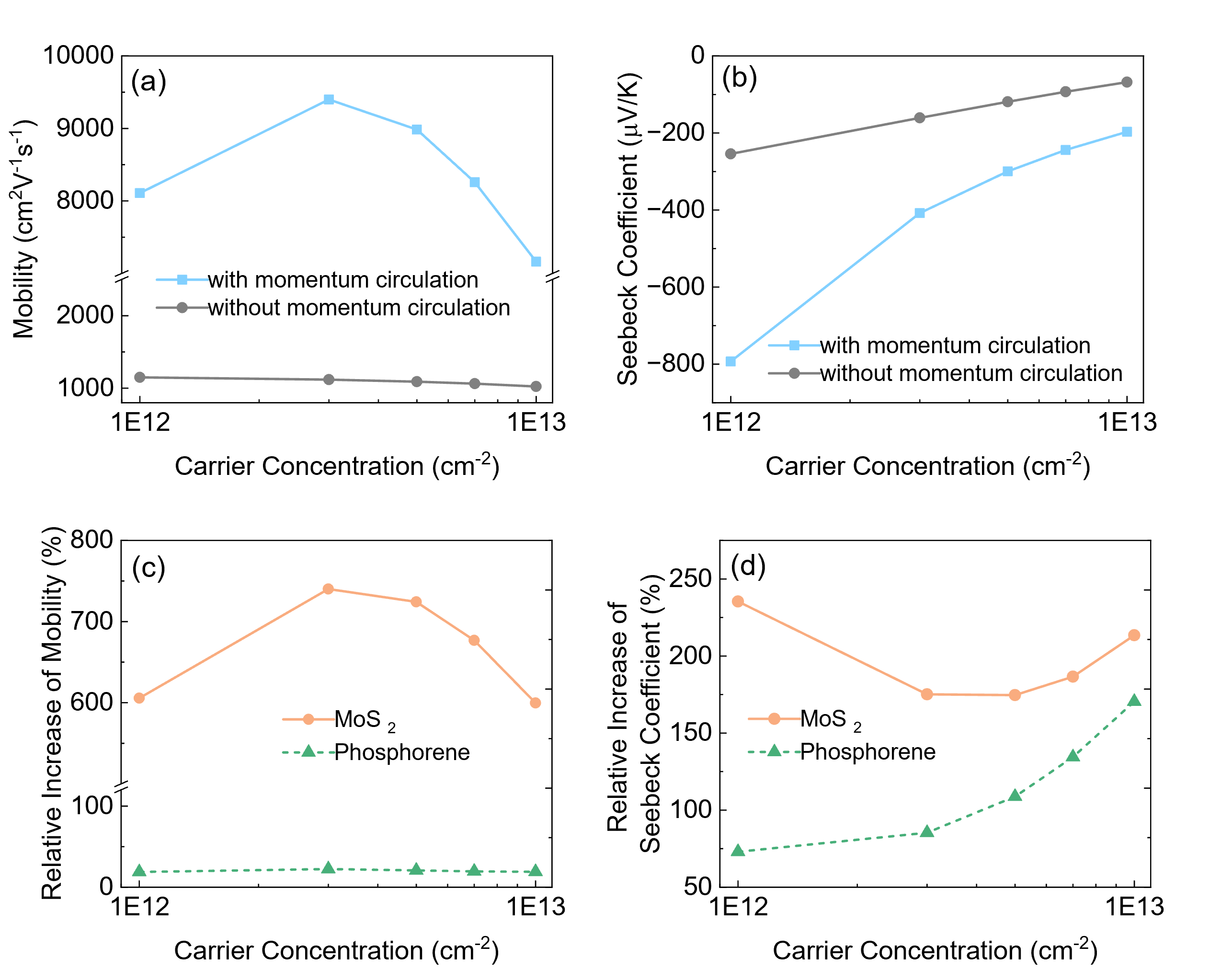}
\caption{\textbf{The calculated transport properties at 100~K.} (a) The calculated mobility of $\rm MoS_2$ with and without the momentum circulation. (b) The calculated Seebeck coefficient of $\rm MoS_2$ with and without the momentum circulation. (c) The relative increase of mobility due to momentum circulation in $\rm MoS_2$ (coupled electron-phonon hydrodynamics) and phosphorene (electron-phonon drag). The electron mobility is significantly increased in the coupled electron-phonon hydrodynamic regime compared with electron-phonon drag. (d) The relative increase of Seebeck coefficient in $\rm MoS_2$ (coupled electron-phonon hydrodynamics) and phosphorene (electron-phonon drag). The increase of the Seebeck coefficient due to coupled electron-phonon hydrodynamics is higher than that due to electron-phonon drag.}
\label{fig:fig4}
\end{figure}

Coupled electron-phonon hydrodynamics in semiconductors has a significant impact on transport properties.
In response to an electric field, charge carriers move through the material along the field direction. In the coupled electron-phonon hydrodynamics regime, electron momentum is less dissipated during transport, leading to an increase in mobility. 
In the Peltier picture, an isothermal electric field can induce a heat flow through interactions between nonequilibrium electrons and phonons, and thus phonons can also contribute to the total Seebeck coefficient~\cite{zhou2015ab}. 
The calculated electron mobility and Seebeck coefficient of $\rm MoS_2$ at 100~K, both with (hydrodynamic regime) and without (non-hydrodynamic regime) the momentum circulation, are shown in Fig.~\ref{fig:fig4}.
When the interactions between nonequilibrium electrons and phonons are not considered, the calculated mobility and Seebeck coefficient agree well with previous literature~\cite{gunst2016first,hippalgaonkar2017high}. 
It is found that by considering the momentum circulation, i.e., in the hydrodynamic transport regime, both electron mobility and Seebeck coefficient increase significantly.
However, the experimentally measured mobility remains notably lower than the predicted values~\cite{baugher2013intrinsic,yu2014towards}. This discrepancy can be attributed to sample quality issues, such as inevitable scattering from short-range defects, traps, and charged vacancies~\cite{kaasbjerg2019electron}, as well as extrinsic factors like contact resistance and substrate interactions. 
Furthermore, to demonstrate the difference between the effect of electron-phonon drag and coupled electron-phonon hydrodynamics, both of which originate from momentum circulation, on transport properties, the relative changes in electron mobility and Seebeck coefficient of phosphorene at 100~K due to the electron-phonon drag effect are also plotted for comparison, which are shown in Fig.~\ref{fig:fig4}(c) and (d).
Because of the momentum circulation, including electron-phonon drag also increases the calculated transport coefficients in phosphorene. 
However, compared with the coupled electron-phonon hydrodynamics in MoS$_2$, in which the total momentum is less dissipated, the increase of mobility and Seebeck coefficient in phosphorene is much less pronounced. 
For example, as shown in Fig.~\ref{fig:fig4}(c), the relative increase of electron mobility in $\rm MoS_2$ is tens of times higher than that in phosphorene.
Similarly, the contribution of phonons to the total Seebeck coefficient is also higher in $\rm MoS_2$ than in phosphorene.
In addition, coupled electron-phonon hydrodynamics also have an impact on thermal conductivity, which is a response to a temperature gradient.
In our previous work, we have demonstrated a significant increase in phonon thermal conductivity in $\rm MoS_2$, while in phosphorene this increase is quite small~\cite{quan2024electron}.
The thermal conductivity increase was attributed to strong momentum circulation resulting from strong electron-phonon interactions, and now this is also recognized as a consequence of coupled electron-phonon hydrodynamics.

As demonstrated above, in the coupled electron-phonon hydrodynamic transport regime, transport properties show a significant increase.
However, it is challenging to detect coupled electron-phonon hydrodynamics by directly measuring the steady-state transport properties given the difficulty to precisely control the sample quality.
%Alternatively, the concept of second sound, arising from phonon hydrodynamics, inspires experimental verification of the coupled electron-phonon hydrodynamics.
%As shown in Fig,~\ref{fig:fig2}, an electric field can drive both electrons and phonons out of equilibrium, which are well described by the displaced equilibrium distributions, showing the hydrodynamic characteristics.
Alternatively, we suggest a transient experiment analogous to the second-sound measurement to detect phonon hydrodynamics. Since electrons and phonons move collectively in the coupled hydrodynamics regime, the phonon second sound initiated by a temperature pulse will be accompanied by an electrical pulse, arriving together on the other side of the sample. This simultaneous detection of a temperature pulse and an electrical pulse could serve as a strong experimental signature of coupled electron-phonon hydrodynamics.

\section{Conclusion}
In summary, we have systematically studied the effect of momentum circulation on electronic and thermal transport properties in 2D semiconductors by solving the fully coupled electron-phonon BTEs.
Despite the strong electron-phonon interactions, which were previously considered the primary mechanism for momentum relaxation in electronic and thermal transport, we find that they lead to less dissipative conduction. 
This low-dissipation conduction is attributed to strong momentum circulation achieved by strong electron-phonon interactions, resulting in a new transport regime\textemdash coupled electron-phonon hydrodynamics.
In this regime, electrons and phonons exhibit drift motion with the same drift velocity.
Furthermore, we distinguish between the electron-phonon drag effect, another result of momentum circulation, and coupled electron-phonon hydrodynamics by highlighting the significant increase in electronic transport properties and the strong influence on thermal conductivity.
We also discussed a potential experimental method for verifying the existence of this coupled electron-phonon hydrodynamics.
Our work provides a new fundamental understanding of electronic and thermal transport properties in 2D semiconductors and points to alternative approaches to achieving low-dissipation charge transport.

\begin{acknowledgments}
This work is based on research supported by the U.S. Air Force Office of Scientific Research under award number FA9550-22-1-0468. Y.Q. also acknowledges support from the NSF Quantum Foundry via the Q-AMASE-i program under award number DMR-1906325 at the University of California, Santa Barbara (UCSB). This work used Stampede2 at Texas Advanced Computing Center (TACC) through allocation MAT200011 from the Advanced Cyberinfrastructure Coordination Ecosystem: Services \& Support (ACCESS) program, which is supported by National Science Foundation grants 2138259, 2138286, 2138307, 2137603, and 2138296. Use was also made of computational facilities purchased with funds from the National Science Foundation (award number CNS-1725797) and administered by the Center for Scientific Computing (CSC) at University of California, Santa Barbara (UCSB). The CSC is supported by the California NanoSystems Institute and the Materials Research Science and Engineering Center (MRSEC; NSF DMR-2308708) at UCSB. 
\end{acknowledgments}

\bibliography{references.bib}% Produces the bibliography via BibTeX.

\end{document}

%\preprint{APS/123-QED}

\title{Supplementary Information: Coupled electron-phonon hydrodynamics in two-dimensional semiconductors} 

\author{Yujie Quan}
\affiliation{Department of Mechanical Engineering, University of California, Santa Barbara, CA 93106, USA}

\author{Bolin Liao}
\email{bliao@ucsb.edu} \affiliation{Department of Mechanical Engineering, University of California, Santa Barbara, CA 93106, USA}

\maketitle

%\tableofcontents
%\renewcommand\linenumberfont{\normalfont\tiny}

%\linenumbers\relax % Commence numbering lines

\section{Methods}
\subsection{Theory}
The coupled electron-phonon hydrodynamics can be captured by solving the coupled electron-phonon BTEs, which was discussed in detail in Ref~\cite{protik2022elphbolt}. 
The collision term in BTEs due to electron-phonon interactions can be written as:
\begin{equation}\label{eqn:2}
    \left\{\begin{aligned}
\left(\dfrac{\partial f_\alpha(\mathbf{k})}{\partial t}\right)_{e-p h} \simeq &-\left[\sum_{\mathbf{k}^{\prime} \beta, \mathbf{q} \lambda} F_{\mathbf{k} \alpha}\left(\mathbf{k}^{\prime} \beta, \mathbf{q} \lambda\right)\right] \cdot \Delta f_{\mathbf{k} \alpha}+\sum_{\mathbf{k}^{\prime} \beta, \mathbf{q} \lambda}\left[F_{\mathbf{k}^{\prime} \beta}(\mathbf{k} \alpha, \mathbf{q} \lambda) \cdot \Delta f_{\mathbf{k}^{\prime} \beta}\right] + \\ &\sum_{\mathbf{k}^{\prime} \beta, \mathbf{q} \lambda}\left[F_{\mathbf{q} \lambda}\left(\mathbf{k} \alpha, \mathbf{k}^{\prime} \beta\right) \cdot \Delta n_{\mathbf{q} \lambda}\right] \\
\left(\dfrac{\partial n_\lambda(\mathbf{q})}{\partial t}\right)_{e-p h} \simeq - &\left[\sum_{\mathbf{k} \alpha, \mathbf{k}^{\prime} \beta} G_{\mathbf{q} \lambda}\left(\mathbf{k} \alpha, \mathbf{k}^{\prime} \beta\right)\right] \cdot \Delta n_{\mathbf{q} \lambda} + \\ & \sum_{\mathbf{k} \alpha, \mathbf{k}^{\prime} \beta}\left[G_{\mathbf{k} \alpha}\left(\mathbf{k}^{\prime} \beta, \mathbf{q} \lambda\right) \cdot \Delta f_{\mathbf{k} \alpha} + G_{\mathbf{k}^{\prime} \beta}(\mathbf{k} \alpha, \mathbf{q} \lambda) \cdot \Delta f_{\mathbf{k}^{\prime} \beta}\right],
\end{aligned}\right.
\end{equation}
in which the last term shows the nonequilibrium phonon (electron) distribution $\Delta n$ ($\Delta f$) driven by nonequilibrium electrons (phonons), leading to possibly coupled electron-phonon hydrodynamics. 
Coefficients $F$ and $G$ only depend on the equilibrium distribution of electrons $f_0$ and phonons $n_0$. 

\subsection{Computational Methods}
DFT calculations were performed using the Quantum ESPRESSO package~\cite{giannozzi2009quantumQE} with the scalar-relativistic optimized norm-conserving Vanderbilt (ONCV) pseudopotentials~\cite{hamann2013optimizedONCV} within the local density approximation (LDA)~\cite{perdew1992accurateLDA}. 
The kinetic energy cutoff for wave functions was set to 80~Ry for all the calculations, and the total electron energy convergence threshold for self-consistency was set to $\rm 1 \times 10^{-10}$~Ry. 
A mesh grid of $18\times 18 \times 1$ and $12\times 12 \times 1$ in the first Brillouin zone was adopted for $\rm MoS_2$ and black phosphorene, respectively.
The crystal lattice is fully relaxed with a force threshold of $\rm 10^{-4}~eV/\textup{\AA}$, with in-plane lattice parameters $\rm a = 3.18 ~\textup{\AA}$ in $\rm MoS_2$, and $\rm a = 3.27 ~\textup{\AA}$, $\rm b = 4.36 ~\textup{\AA}$ in phosphorene, both of which were in excellent agreement with the experimental values~\cite{vancso2016intrinsic,jain2015strongly}.
To avoid interactions between layers, a vacuum layer of $\rm 30~ \textup{\AA}$ and $21~\textup{\AA}$ was introduced along the axis perpendicular to the $\rm MoS_2$ and black phosphorene monolayer, respectively.
The second-order interatomic force constants (IFCs) were calculated using DFPT \cite{baroni2001phononsDFPT} as implemented in Quantum ESPRESSO, with a $6 \times 6 \times 1$ q-point grid for $\rm MoS_2$ and phosphorene.
The third-order IFCs were computed using a $6 \times 6 \times 1$ supercell for both $\rm MoS_2$ and phosphorene with the finite displacement method~\cite{li2014shengbte}, taking up to the fifth nearest neighbors into consideration.
The electron-phonon matrix elements were calculated using the EPW package~\cite{ponce2016epw}, with a $18 \times 18 \times 1$ k-point grid and a $6 \times 6 \times 1$ q-point grid for $\rm MoS_2$, and a $12 \times 12 \times 1$ k-point grid and a $6 \times 6 \times 1$ q-point grid for phosphorene.
The electron-phonon matrix elements in Wannier space as output of the EPW package were used as input for transport calculations with Elphbolt~\cite{protik2022elphbolt}, where real-space Wannier quantities were transformed into Bloch representations on a fine $150 \times 150 \times 1$ q-point grid and a $300 \times 300 \times 1$ k-point grid for $\rm MoS_2$ and a fine $100 \times 100 \times 1$ q-point grid and a $300 \times 300 \times 1$ k-point grid for phosphorene.
The expression of Fr\"ohlich interactions in 2D $\rm MoS_2$ was adapted from Ref.~\cite{sio2022unified} with the simplest approximation, which shows excellent agreement with the 2D Coulomb truncation method~\cite{sohier2016two}.
Phonon-isotope scatterings and electron-charged-impurity scatterings were also considered in the transport calculations, in which the Tamura~\cite{tamura1983isotope} and Brooks-Herring models~\cite{brooks1955theory} were adopted, respectively.

\section{Response to a unit Electric Field}
\begin{figure}[!htb]
\includegraphics[scale=0.8]{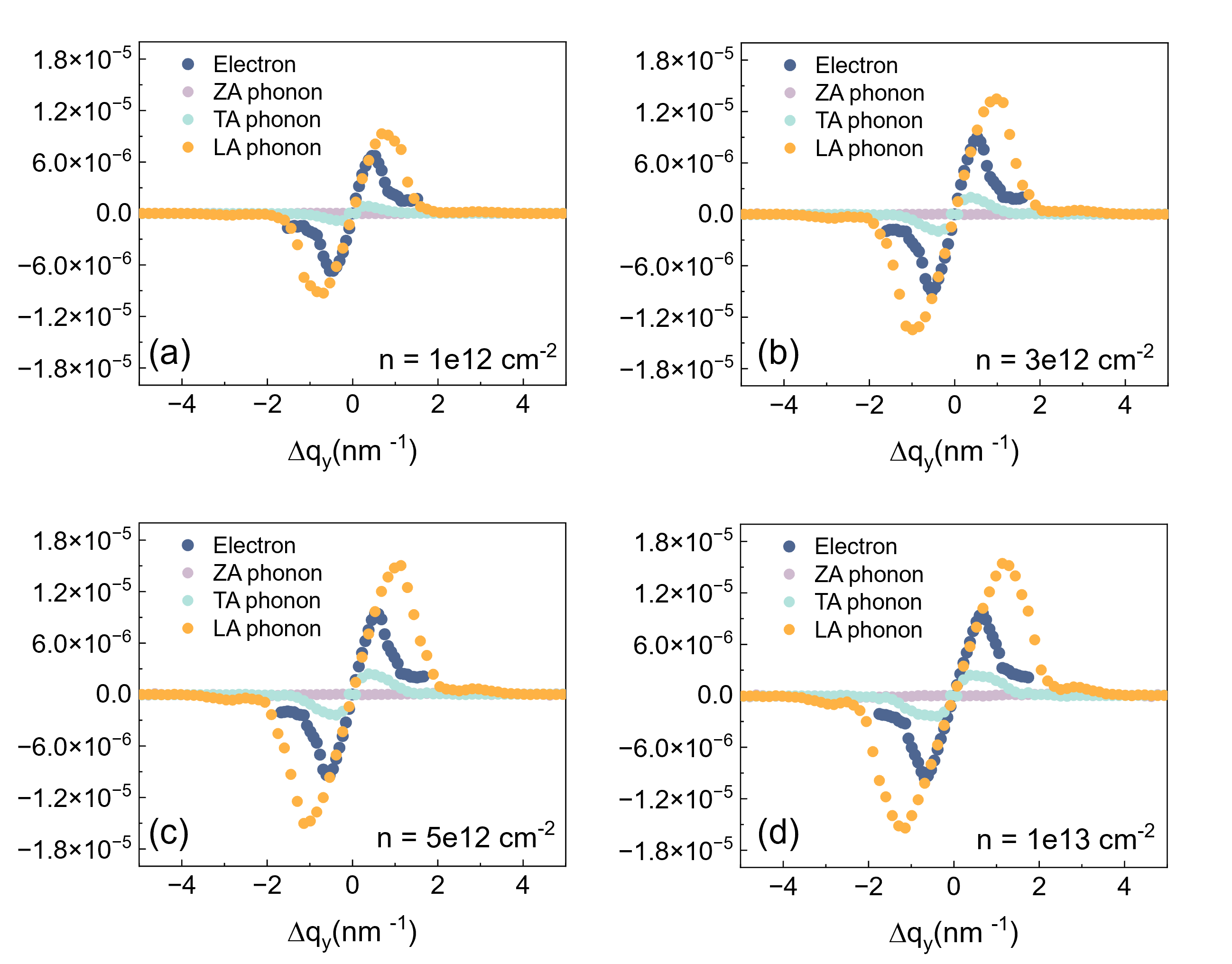}
\caption{The normalized deviation of $\rm MoS_2$ at 150~K in response to a unit electric field.} 
\label{fig:figS1}
\end{figure}

\begin{figure}[!htb]
\includegraphics[scale=0.8]{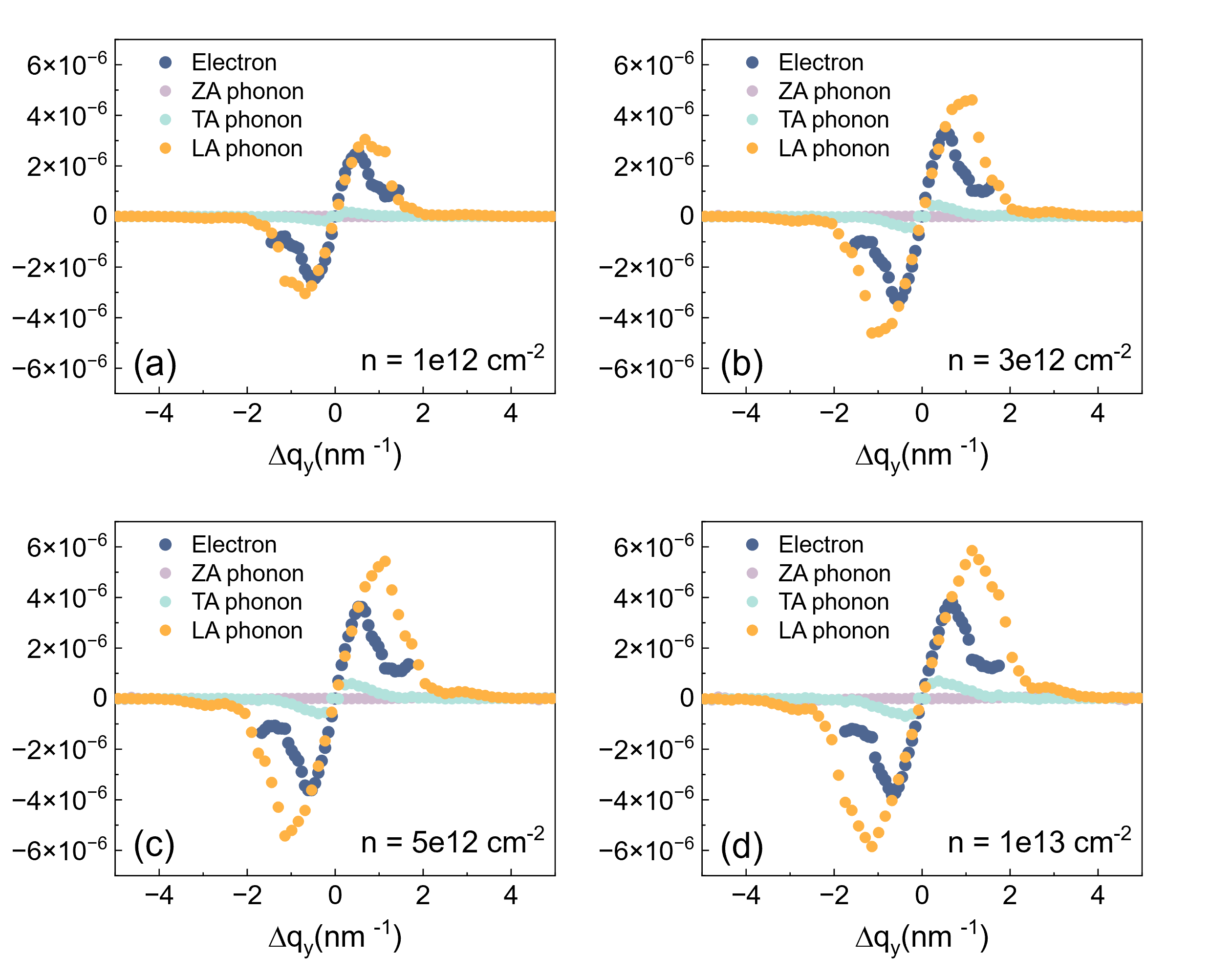}
\caption{The normalized deviation of $\rm MoS_2$ at 200~K in response to a unit electric field.} 
\label{fig:figS2}
\end{figure}

\begin{figure}[!htb]
\includegraphics[scale=0.8]{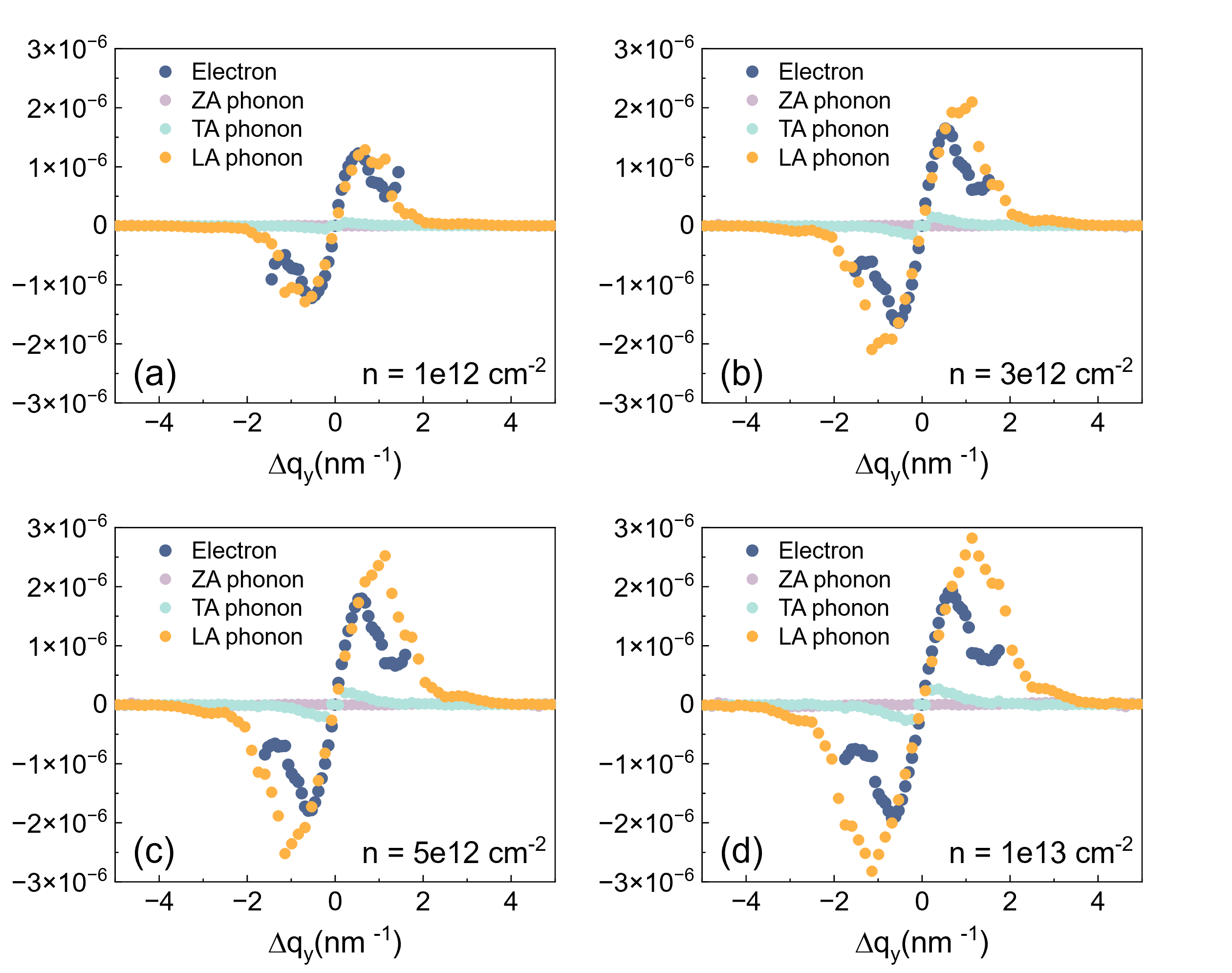}
\caption{The normalized deviation of $\rm MoS_2$ at 250~K in response to a unit electric field.} 
\label{fig:figS3}
\end{figure}

\begin{figure}[!htb]
\includegraphics[scale=0.8]{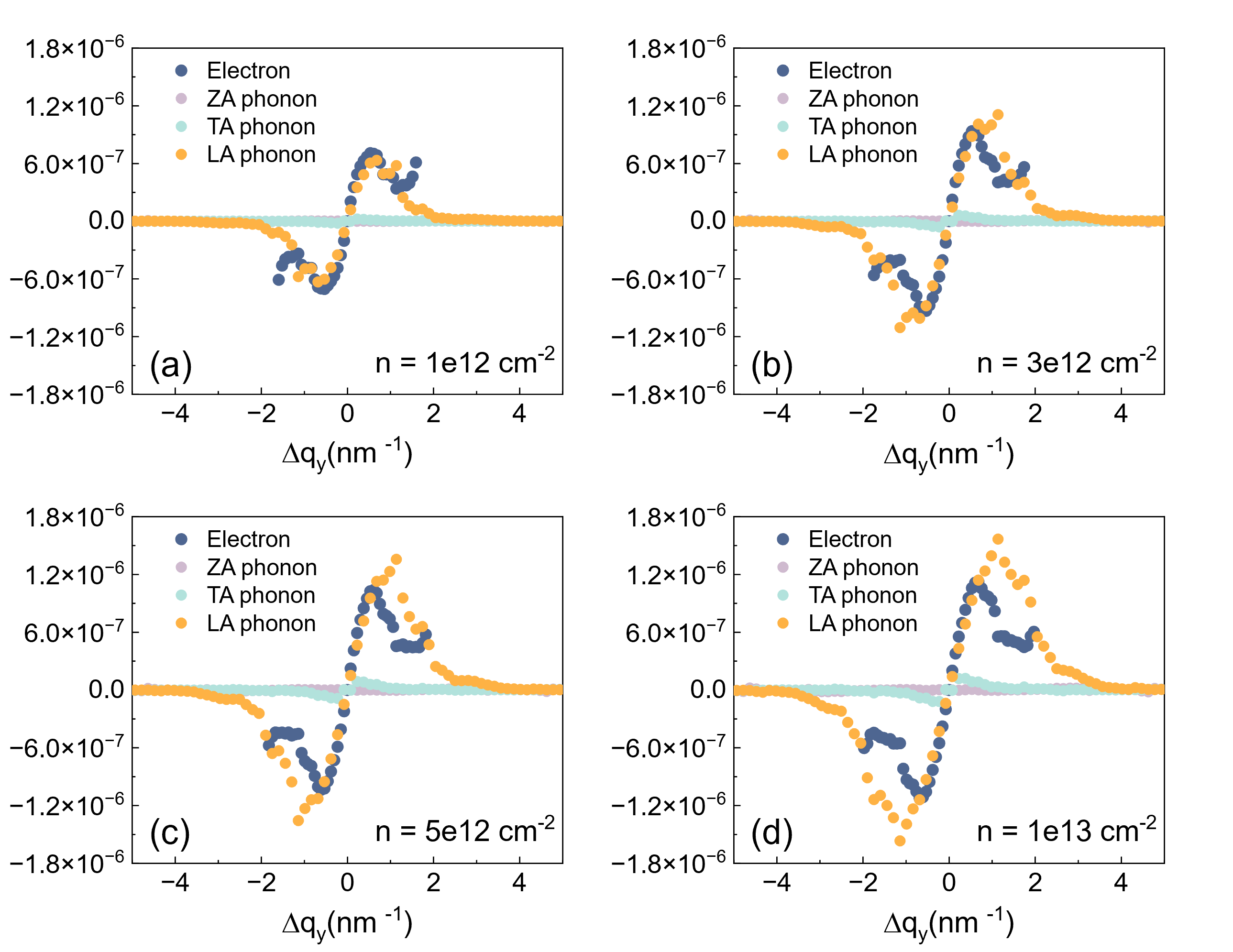}
\caption{The normalized deviation of $\rm MoS_2$ at 300~K in response to a unit electric field.} 
\label{fig:figS4}
\end{figure}

The normalized deviations of electron and phonon distribution of $\rm MoS_2$ in response to a unit electric field at higher temperatures are shown in Fig.~\ref{fig:figS1} - \ref{fig:figS4}.

\begin{figure}[!htb]
\includegraphics[scale=0.4]{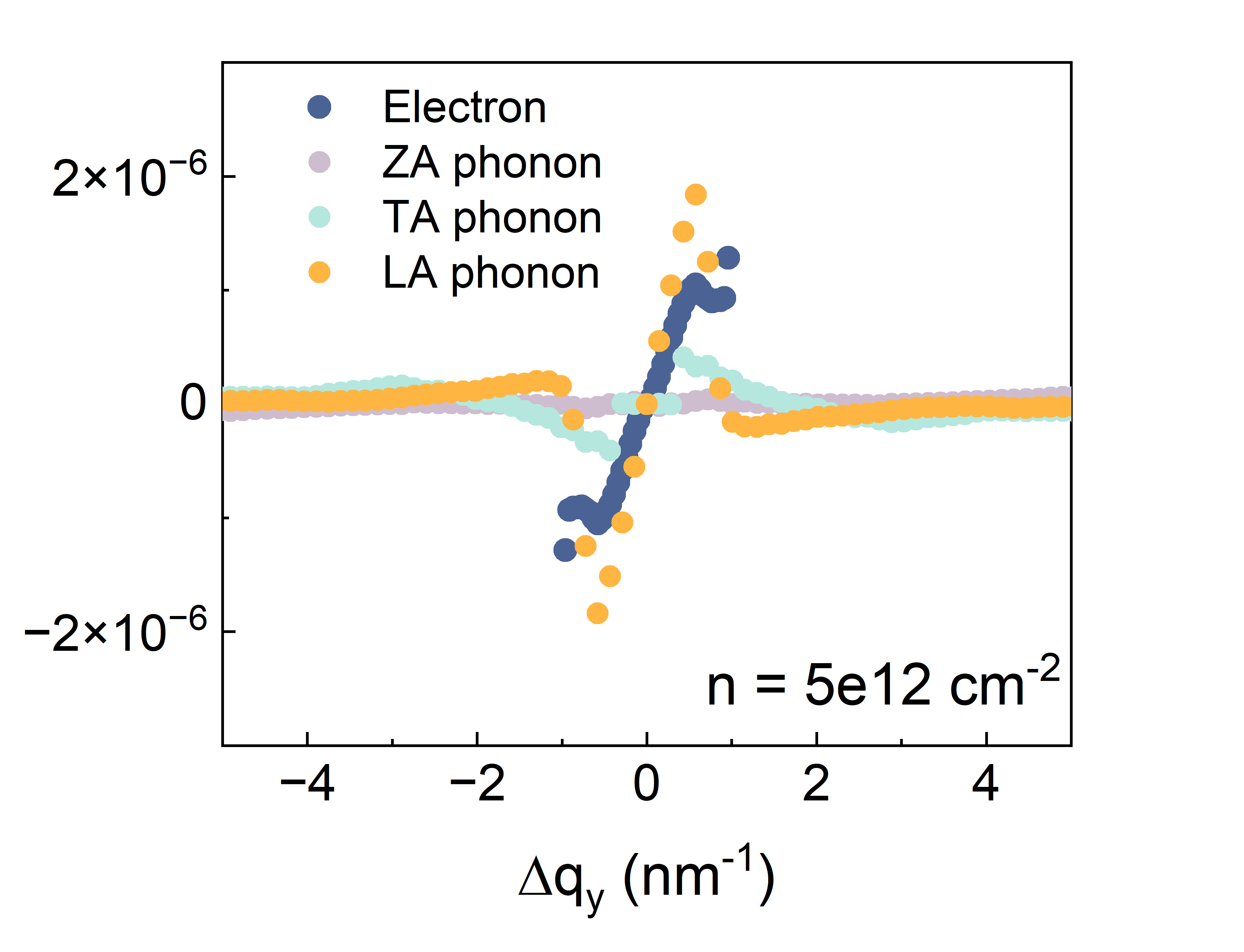}
\caption{The normalized deviation of black phosphorene at 100~K in response to a unit electric field. Unlike $\rm MoS_2$, there is no coupled electron-phonon hydrodynamics feature.} 
\label{fig:figS5}
\end{figure}

The normalized deviations of electron and phonon distribution of black phosphorene in response to a unit electric field at 100~K are shown in Fig.~\ref{fig:figS5}. Unlike $\rm MoS_2$, electrons and phonons in black phosphorene do not exhibit joint drift motion, indicating the absence of coupled electron-phonon hydrodynamics. 

\section{Response to a unit temperature gradient}
\begin{figure}[!htb]
\includegraphics[scale=0.58]{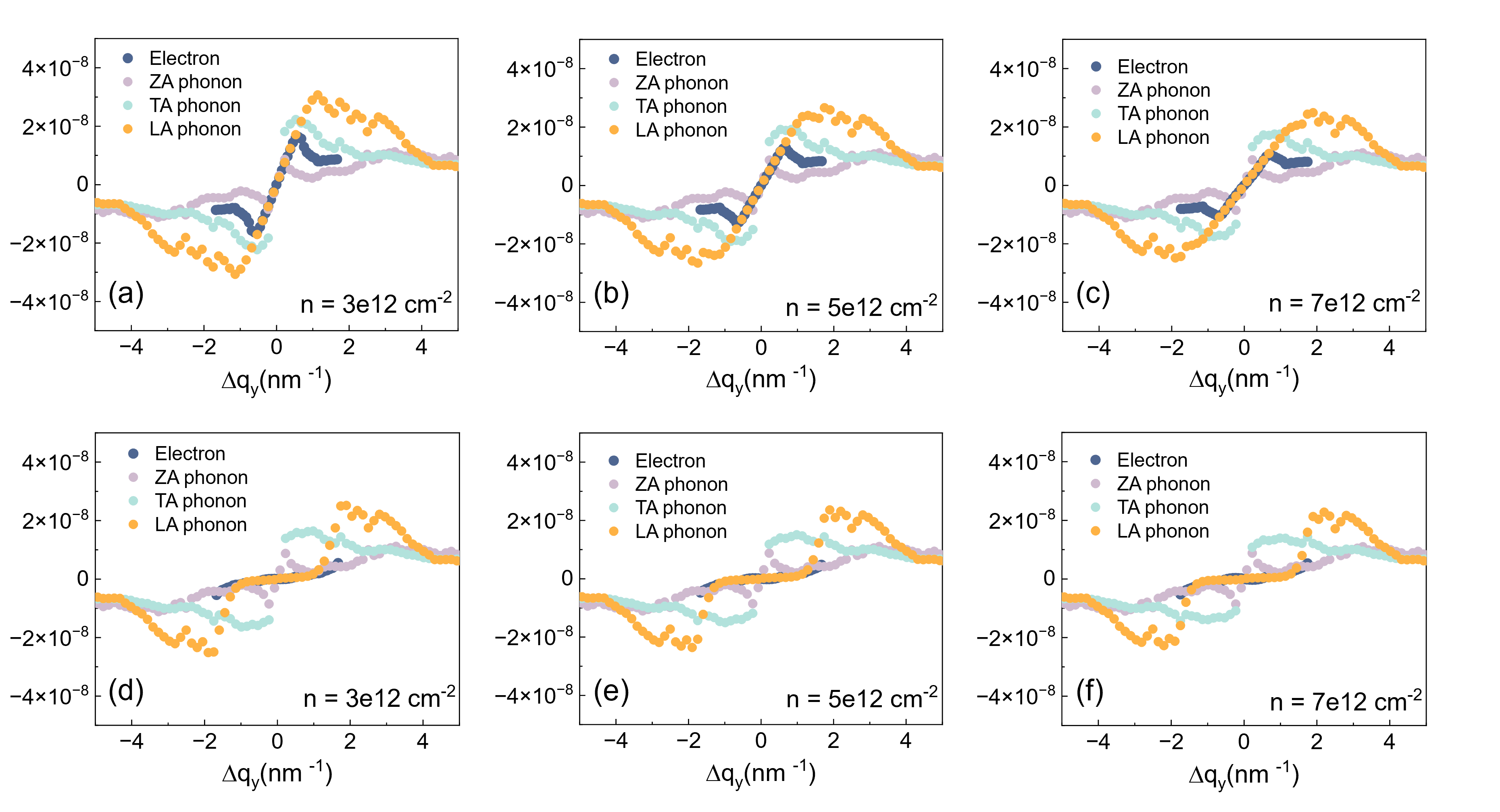}
\caption{The normalized deviation of $\rm MoS_2$ at 100~K in response to a unit temperature gradient.} 
\label{fig:fig100K}
\end{figure}

\begin{figure}[!htb]
\includegraphics[scale=0.58]{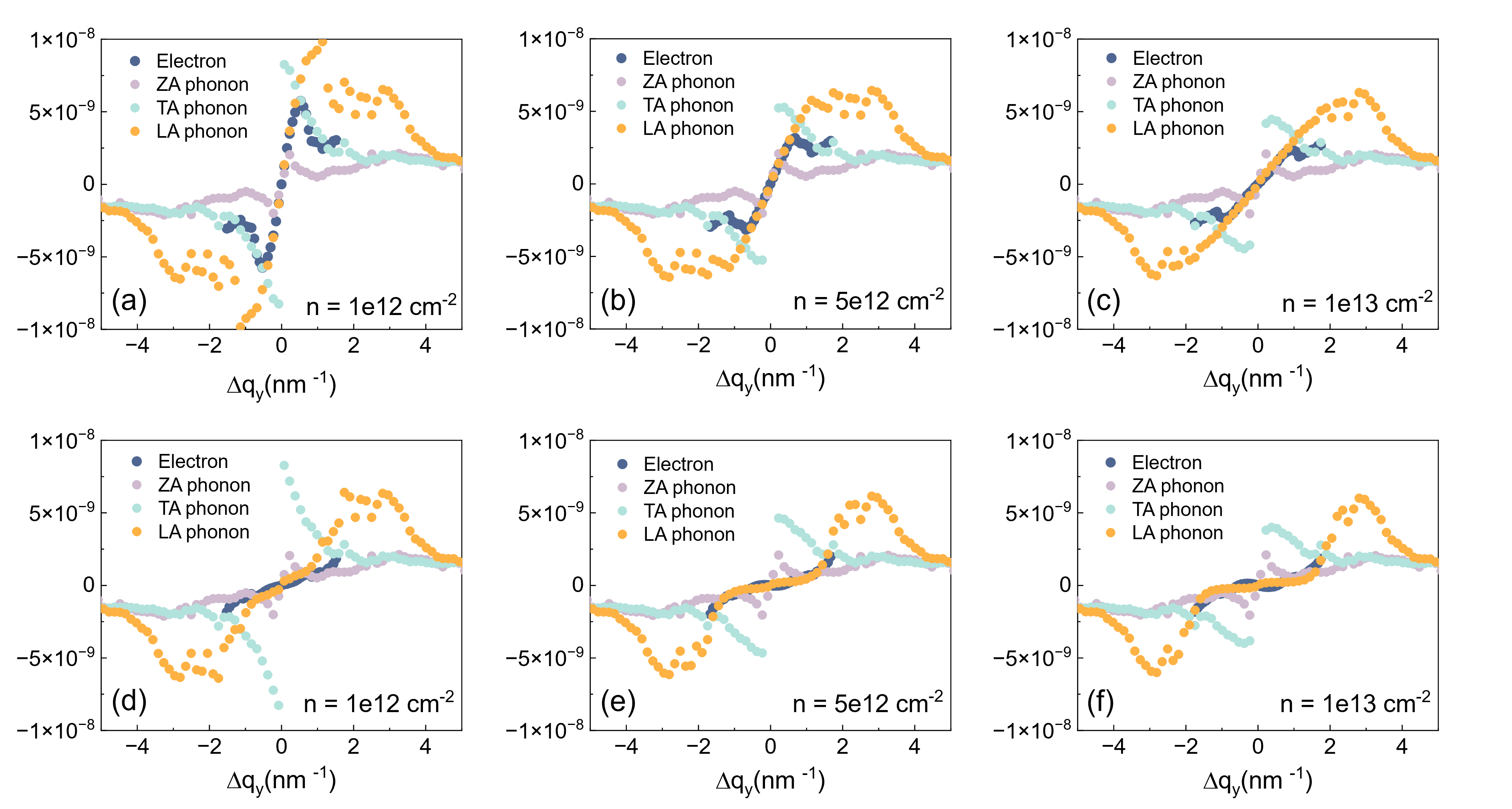}
\caption{The normalized deviation of $\rm MoS_2$ at 150~K in response to a unit temperature gradient.} 
\label{fig:figS6}
\end{figure}

\begin{figure}[!htb]
\includegraphics[scale=0.58]{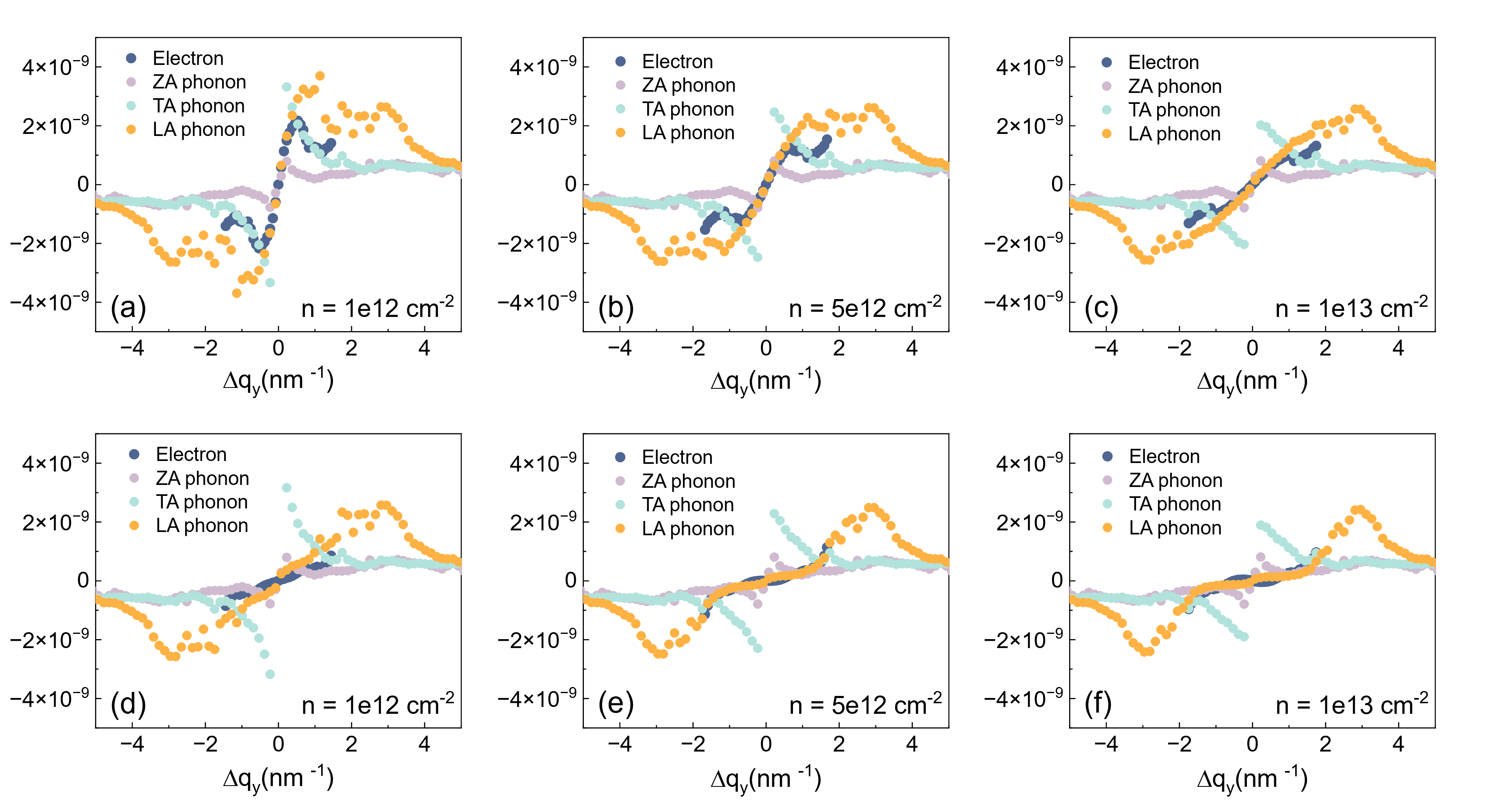}
\caption{The normalized deviation of $\rm MoS_2$ at 200~K in response to a unit temperature gradient.} 
\label{fig:figS7}
\end{figure}

\begin{figure}[!htb]
\includegraphics[scale=0.58]{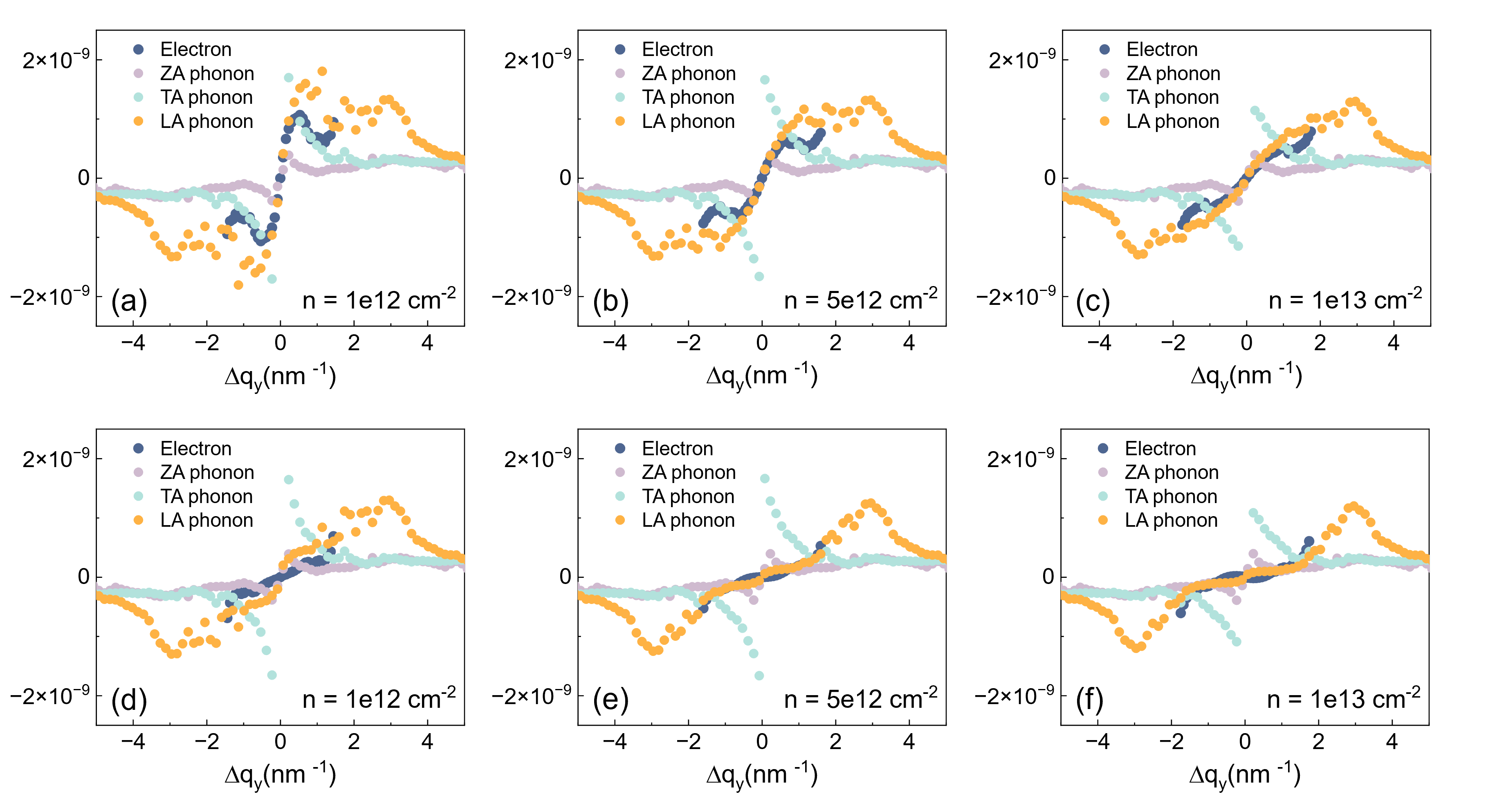}
\caption{The normalized deviation of $\rm MoS_2$ at 250~K in response to a unit temperature gradient.} 
\label{fig:figS8}
\end{figure}

\begin{figure}[!htb]
\includegraphics[scale=0.58]{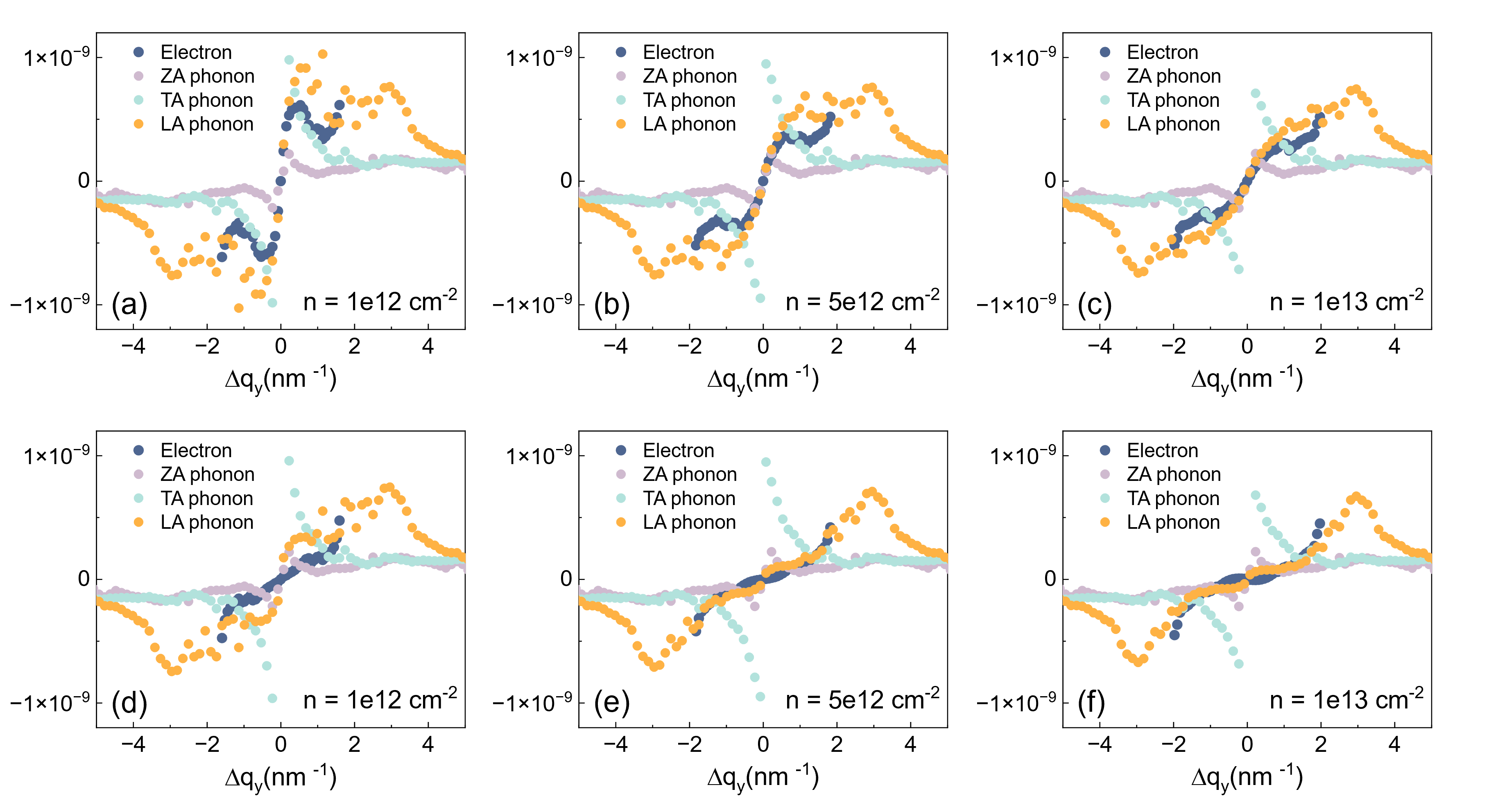}
\caption{The normalized deviation of $\rm MoS_2$ at 300~K in response to a unit temperature gradient.} 
\label{fig:figS9}
\end{figure}

The normalized deviations of electron and phonon distribution of $\rm MoS_2$ in response to a unit temperature gradient at 100~K and higher temperatures are shown in Fig~\ref{fig:fig100K} - \ref{fig:figS9}. The top panels show the normalized deviation considering the interactions between nonequilibrium electrons and phonons, whereas the bottom panels show the normalized deviation without considering these interactions.

\begin{figure}[!htb]
\includegraphics[scale=0.38]{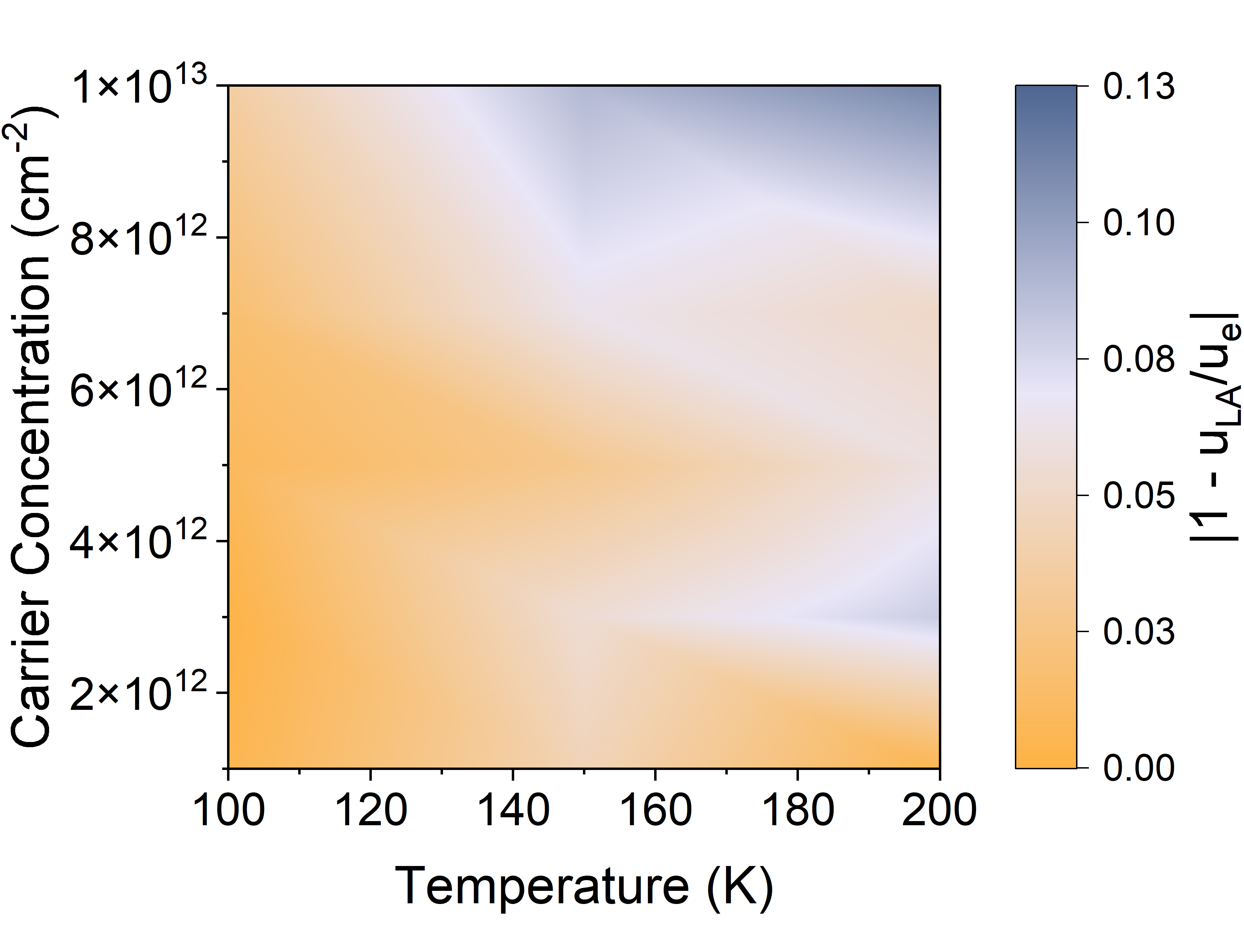}
\caption{The deviation of the drift velocity ratio between LA phonons and electrons from unity in response to a temperature gradient as a function of temperature and carrier concentration.} 
\label{fig:figS10}
\end{figure}

\begin{figure}[!htb]
\includegraphics[scale=0.8]{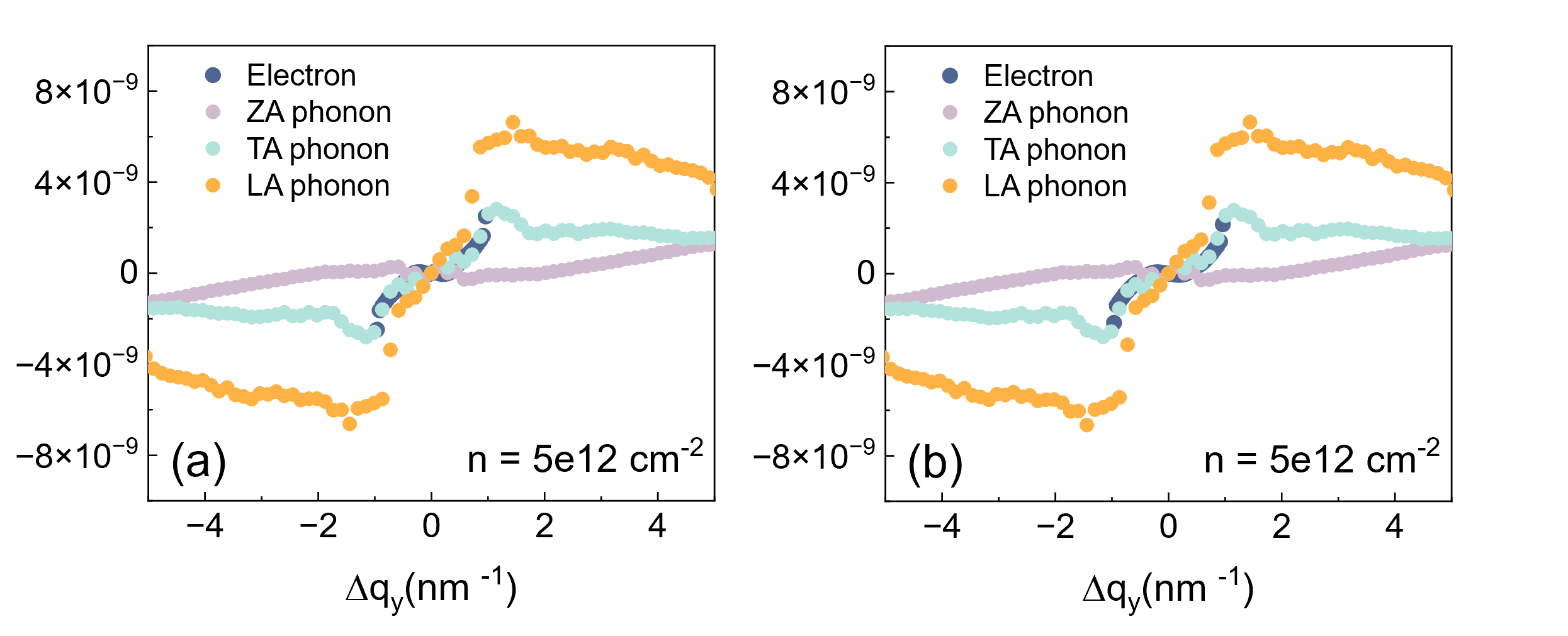}
\caption{The normalized deviation of black phosphorene at 100~K in response to a unit temperature gradient (a) with and (b) without considering the interactions between nonequilibrium electrons and phonons. } 
\label{fig:figS11}
\end{figure}

To compare, the normalized deviation of black phosphorene in response to a unit temperature gradient at 100~K is shown in Fig.~\ref{fig:figS11}. Panel (a) shows the normalized deviation considering the interactions between nonequilibrium electrons and phonons, whereas panel (b) shows the normalized deviation without considering these interactions. Different from $\rm MoS_2$, in which the distribution functions of both electrons and phonons change significantly and electrons and phonons enter a coupled hydrodynamic regime due to the interactions between nonequilibrium electrons and phonons, the distribution functions in black phosphorene change very little. 

\clearpage
\bibliography{references.bib}% Produces the bibliography via BibTeX.